\documentstyle[epsfig]{aa}

\begin{document}
\title{Dust emission in early-type galaxies: The mid-infrared view.
\thanks{Based on observations with ISO, an ESA project with instruments
funded by ESA Member States (especially the PI countries: France, Germany,
the Netherlands and the United Kingdom) and with the participation
of ISAS and NASA.}}
\author{E.M. Xilouris, S.C. Madden, F. Galliano, 
L. Vigroux, M. Sauvage } 
\offprints{S.C. Madden, \email{smadden@cea.fr}}
\institute{
DAPNIA/Service d'Astrophysique, CEA/Saclay, 91191 Gif--sur--Yvette
Cedex, France}

\date{Received .... / Accepted ....} 

\abstract{We present mid-infrared (MIR) maps for a sample of 18 early-type galaxies observed
at 4.5, 6.7 and 15 $\mu m$ with the ISOCAM 
instrument on board the ISO satellite with a 6'' spatial resolution. 
We model the Spectral Energy Distribution (SED)
of these galaxies using
the stellar evolutionary synthesis model P\'EGASE and we derive the MIR
excess over the stellar component.
We then explore the nature of this excess in terms of dust and Polycyclic
Aromatic Hydrocarbon molecules (PAHs). 
We find that out of 18 galaxies, 10 show excess at 6.7 $\mu m$ (due to the
presence of PAH features) and 14 show excess at 15 $\mu m$ (due to the presence
of warm dust). In two galaxies, where a more complete wavelength coverage
exists, an excess around 9.7 $\mu m$ is seen (presumably due to silicate dust emission),
while two other galaxies are totally devoid of dust.
We also examine the morphology of the galaxies in these wavelengths by plotting the
azimuthally averaged radial profiles as well as the MIR color profiles. 
We find that for the majority of the galaxies the 4.5 $\mu m$ emission is well 
described by a de Vaucouleurs profile. The 6.7 $\mu m$ and 15 $\mu m$ emission
is smoothly distributed throughout the galaxy while only a few galaxies show
MIR emission which is more concentrated close to the center. Two dwarf 
galaxies in our sample show patchy distributions of the MIR emission while
two other galaxies show edge-on disks.
With color-color
diagrams we delineate the regions occupied by late-type and early-type galaxies.
Finally we show that the MIR excess found in strong radio galaxies like NGC 4486 (M87)
can be explained by synchrotron emission.

\keywords{ISM: dust, extinction -- Galaxies: elliptical and lenticular, cD -- Galaxies: ISM -- 
Infrared: galaxies -- Infrared: ISM}}
\titlerunning{Dust emission in early-type galaxies}
\authorrunning{Xilouris et al.}
\maketitle

%%%%%%%%%%%%%%%%%%%%%%%%%%%%%%%%%%%%%%%%%%%%%%%%%%   
\section{Introduction}
%%%%%%%%%%%%%%%%%%%%%%%%%%%%%%%%%%%%%%%%%%%%%%%%%%

One of the discoveries of {\it Infrared Astronomical Satellite} (IRAS) was that
a significant fraction of early-type galaxies (ellipticals and lenticulars)
contain non negligible amounts of dust (Jura et al. 1987, Knapp et al. 1989)
although the detection rate drops significantly when active
galactic nuclei (AGN), peculiar systems and background contamination are excluded
(Bregman et al. 1998).
This was a surprising
result since early-type galaxies were thought to be relatively devoid of interstellar
medium (ISM). Recent high resolution images from the {\it Hubble Space Telescope} (HST)
verify that about 78\% of the early-type galaxies contain nuclear dust 
(van Dokkkum \& Franx 1995).

These discoveries have lead a number of groups over the 
last few years to examine the physical conditions  and nature of the ISM in 
early-type galaxies. The {\it Infrared Space Observatory} (ISO), with greater 
sensitivity and better spatial resolution compared to IRAS, has
proven to be a very useful tool to probe the dust content in these
galaxies. 
Mid-infrared (MIR) observations made by ISO have shown that even though the evolved stellar population
can be a dominant contributor in this part of the spectrum (Boselli et al. 1998),
dust in the form of either
small hot grains and/or Polycyclic Aromatic Hydrocarbon molecules (PAHs)
can also be present in some cases 
(Madden et al. 1999, Malhotra et al. 2000, Athey et al. 2002, 
Ferrari et al. 2002).

One interesting question pertains to the origin of dust in early-type galaxies.
Possible scenarios for the origin of the dust in early-type galaxies include:
1) merging events resulting in galaxies donating large amounts of dust and
gas, often seen in dust patches and dust lanes 2) cooling flow condensations
(e.g. Fabian et al. 1991) often accounting for dust centrally concentrated
and 3) mass loss from late-type stars (Knapp et al. 1992) which would result 
in dust more widely distributed throughout the galaxy.

The evolution of the dust grains in the hot X-ray environment of some elliptical
galaxies is also of great interest. It is remarkable that despite the fact 
that thermal sputtering destroys the grains, these galaxies can harbor a
substantial amount of dust (Tsai \& Mathews, 1995).

In this paper we study the MIR emission of 18 early-type galaxies observed with 
the ISOCAM instrument on board the ISO satellite for which we have
obtained broad band imaging at 4.5, 6.7 and 15 $\mu m$ and which are well studied
at many wavelengths.
For each galaxy we perform detailed modelling of the Spectral Energy Distribution (SED)
using the P\'EGASE stellar evolutionary synthesis model  (Fioc \& Rocca-Volmerange 1997).
We use this model to fit the available data from the ultra-violet (UV) to near-infrared 
(NIR) part of the spectrum. Extrapolating the modelled SED to the MIR wavelengths
we then determine the MIR excess over the old stellar component. 
We examine the MIR colors
as a function of the radius of the galaxy and we present diagnostic color-color diagrams
of MIR and far-infrared (FIR) colors in order to delineate regions occupied by
different types of galaxies. We also examine the morphology of these galaxies 
using their azimuthally averaged radial profiles. 

The paper is structured as follows. In Section 2, we describe the galaxy
sample and the observations. In Section 3, we describe our data reduction 
procedure.
In Section 4, we give a brief description of the MIR extragalactic
spectra and what features can be present in the ISOCAM bands.
The results of this study are presented in Section 5
and a discussion follows in Section 6. We summarize the main
points of this study in Section 7.

%%%%%%%%%%%%%%%%%%%%%%%%%%%%%%%%%%%%%%%%%%%%%%%%%%
\section{Observations}
%%%%%%%%%%%%%%%%%%%%%%%%%%%%%%%%%%%%%%%%%%%%%%%%%%
The observations were carried out with the ISOCAM $32 \times 32$ LW imaging array (Cesarsky et al. 1996)
on board the ISO satellite (Kessler et al. 1996). 
The galaxy images
were constructed in raster mode with individual frames slightly 
shifted (usually $\sim 10''$) and combined into a $2\times 2$ 
raster configuration.
%%%%%%%%%%%%%%%%%%%%% TABLE 1 %%%%%%%%%%%%%%%%%%%%
\begin{table}
\begin{minipage}{8.7truecm}
\begin{center}
\caption{Observational information of the galaxies.} 
\begin{tabular}{lccc} \hline \hline
NAME & ISO-TDT\footnote{The Target Dedicated Time (TDT) number of the {\it ISO}
observations as it appears in the {\it ISO} archive.} &Hubble type & Distance  \\
     &           &     & (Mpc)    \\ \hline
NGC 185 \footnote{6'' observations are only available.}  & 40001320 &dE5  & 0.62  \\
NGC 205$^b$  & 39902919 & dE5 & 0.72  \\
NGC 1052 \footnote{6'' observations for the 4.5, 6.7 and 15 $\mu m$ as well as 
3'' observations for the 6.7 and 15 $\mu m$ are available.}
& 79601801 & E4 & 17.70  \\
NGC 1316$^c$  & 58800702 &S0 & 18.11\\
NGC 1399 $^{c,}$\footnote{Additional 6'' observations at 6.0, 6.8, 7.7, 9.6, 11.3 and 14.9 $\mu m$
are also available.}  & 85201724 & E1 & 18.11  \\
NGC 2300$^c$  & 72601523 &SA0 & 27.67  \\
NGC 3928$^c$  & 15301004& E1 & 17.00  \\
NGC 4278$^c$  & 22900605& Sa &16.22  \\
NGC 4374$^c$  & 23502406 &E0 & 15.92  \\
NGC 4473$^c$  & 23800407 &E1 & 16.14  \\
NGC 4486$^c$  & 23800808 &E5 & 15.92 \\
NGC 4649$^c$  & 24101209 &E2 & 15.92  \\
NGC 5018$^c$  & 40000210 &E2 & 30.20 \\
NGC 5102$^{c,d}$ & 08902411& SA0  & 4.16  \\
NGC 5173$^c$  & 35400412& E0 & 34.99 \\
NGC 5266$^c$  & 08100313 &SA0 & 60.00 \\
NGC 5363$^c$  & 22500814& S0 & 15.79 \\
NGC 5866$^c$  & 08001315& SA0 & 13.18  \\
\end{tabular}
\end{center}
\end{minipage}
\end{table}
%%%%%%%%%%%%%%%%%%%%%%%%%%%%%%%%%%%%%%%%%%%%%%%%%%%%%%%%%%%%%%%%%%%%%%
Each galaxy was observed in three broadband
filters, LW1 centered at $\lambda_c= 4.5 \mu m$ ($\Delta \lambda= 4 - 5 \mu m$),
LW2 centered at $\lambda_c= 6.7 \mu m$ ($\Delta \lambda= 5 - 8.5 \mu m$) and LW3
centered at $\lambda_c= 15 \mu m$ ($\Delta \lambda= 12 - 18 \mu m$). We shall hereafter
refer to them by their central wavelength. The
observations were performed using a $6''$ pixel size with a beam FWHM of
$\sim$ 5.2'', 6.0'', 6.5'' for the 4.5, 6.7, 15 $\mu m$ observations respectively.
At the mean distance to our sample of galaxies ($\sim 20$ Mpc) the pixel
scale corresponds to $\sim 0.6$ kpc.
For the closest galaxies in our sample (the two dwarf elliptical galaxies
NGC 185 and NGC 205) the pixel scale is $\sim 18$ pc.
Additional observations for NGC 1399 and
for NGC 5102 in the
LW4 ($\lambda_c= 6.0 , \Delta \lambda= 5.5 - 6.5$),
LW5 ($\lambda_c= 6.8, \Delta \lambda= 6.5 - 7.0$),
LW6 ($\lambda_c= 7.7, \Delta \lambda= 7.0 - 8.5$),
LW7 ($\lambda_c= 9.6, \Delta \lambda= 8.5 - 10.7$),
LW8 ($\lambda_c= 11.3, \Delta \lambda= 10.7 - 12.0$),
LW9 ($\lambda_c= 14.9, \Delta \lambda= 14.0 - 16.0$)
and with a $6''$ pixel size already presented in
Athey et al. (2001) are re-analyzed and presented here. Table 1
gives basic data about the galaxies in the sample. Distances
are from O' Sullivan et al. (2001).

%%%%%%%%%%%%%%%%%%%%%%%%%%%%%%%%%%%%%%%%%%%%%%%%%%
\section{Data reduction}
%%%%%%%%%%%%%%%%%%%%%%%%%%%%%%%%%%%%%%%%%%%%%%%%%%
Data reduction was performed using the CAM Interactive
Reduction package (CIR; Chanial 2003). The analysis (which is similar to that described
in great detail in Madden et al. 2003) proceeded as follows.
First, the dark current subtraction was done using a method based on the Biviano
et al. (1998) model. Then, the borders of the detector,
which are insufficiently illuminated,
were masked out in order to allow for proper photometric analysis. 
The cosmic ray removal was done in two 
steps. In the first step, an automatic method based on the
Multiresolution Median Transform (Starck et al., 1999)
was used to correct for the major cosmic ray events. Then a very
careful visual inspection of the time history of each pixel
was made and a mask was applied to pixels that showed 
deviations from the mean value for a given configuration
(i.e. for the same pointing, the same pixel size and the same filter).
Transient corrections were made in order
to account for memory effects using the Fouks-Schubert method (Coulais
\& Abergel, 2000).
After averaging all frames of the same 
configuration (same filter and pixel size) we corrected for the flat-fielding
effects.
Because of the large extent of our objects
we used two different
methods to do the flat-field correction. In most cases the flat-field
response was calculated by averaging the values of those pixels that
were illuminated by the background emission (after masking the source)
and taking interpolated values from the neighboring pixels and the
default calibration flat-field for the masked pixels. 
For observations where the object was quite extended (i.e. little
background is present in the maps) the raw calibration flats 
were used (see Roussel et al. 2001, for a more detailed
description).
After flat-fielding the images, the final map was
constructed by projecting together all the individual images 
of the same configurations. Then, the conversion from electronic
units to flux density units was made, using the in-flight
calibration data base. 
The sky background was determined in the galaxy frames by calculating
the median value of the sky pixels (after masking the source). 
In some cases (all the 4.5 $\mu m$ observations and the NGC 4486
and NGC 5363 6.7 $\mu m$ observations) the field of view was smaller than
the size of the galaxy and the background was not present in the images. In those cases 
we created azimuthally averaged profiles of the galaxies and fitted these 
with a de Vaucouleurs profile (de Vaucouleurs 1953) allowing for the background value to 
vary as a free parameter. An example is shown in Fig. 1 with the 
azimuthally averaged radial profile of the 4.5 $\mu m$ observations of NGC 1399
(points) fitted with a de Vaucouleurs profile (dot-dashed line)
and a constant background (dashed line). The solid line gives the best fit to
the data (with the innermost part of the profile not fitted).

%%%%%%%%%%%%%%%%%%%%%%%%%%%%%%%%%%%%%%%%%%%%%%%%%%
\section{Mid-IR extragalactic spectra}
%%%%%%%%%%%%%%%%%%%%%%%%%%%%%%%%%%%%%%%%%%%%%%%%%%

The MIR wavelengths trace a variety of environments such as ionized gas
(traced by the ionic lines), photodissociation regions (traced by the PAHs)
as well as emission from stellar photospheres and warm dust found throughout
a galaxy.
More specifically, the three ISOCAM filters used in this survey (4.5, 6.7 and
15 $\mu m$) can be used to trace the following possible components:

\begin{itemize}
\item 4.5 $\mu m$ (LW1): The Rayleigh-Jeans tail of the evolved stellar population.

\item 6.7 $\mu m$ (LW2): The PAH features at 6.3, 7.7 and 8.6 $\mu m$, if present
and excited, can dominate the emission in this filter. These bands are proposed to be
of carbonaceous origin with sizes from 0.4 nm to 1.2 nm, stochastically heated to high temperatures
(L\'{e}ger \& Puget 1984; Allamandola et al. 1990; D\'{e}sert et al., 1990).
Low level continuum emission from the evolved stellar population may also be present.
In the special cases of AGNs, very warm circumnuclear dust can also contribute
to this band.

\item 15 $\mu m$ (LW3): Continuum emission in this band is seen in HII regions and
IR luminous galaxies, thought to originate in very small grains (VSGs; with sizes from
1.2 nm to 15 nm) which are transiently
heated to high temperatures, i.e. 100 to 200 K (D\'{e}sert et al., 1990). The PAHs at 11.3 and
12.8 $\mu m$ can also be present in this filter as well as
emission from nebular lines, e.g. NeII, NeIII. Non-thermal emission associated with
AGNs and central radio sources can also dominate the emission in this band.
\end{itemize}

\noindent
If no dust were present, the 6.7 and 15 $\mu m$ bands would only contain the
diminishing emission of the old stellar population.

%%%%%%%%%%%%%%%%%%%%%%%%%%%%%%%%%%%%%%%%%%%%%%%%%%
\section{Results}
%%%%%%%%%%%%%%%%%%%%%%%%%%%%%%%%%%%%%%%%%%%%%%%%%%

\subsection{Mid-IR maps}
For each galaxy, we present the ISOCAM $4.5, 6.7$ and $15 \mu m$ observations in 
Figs. A.1 - A.9 of Appendix A.
The background image overlaid with the $4.5 \mu m$ contours 
in the top panel, is an optical map from the Digitized
Sky Survey (except for NGC 4486 (M87) where a WFPC-2 HST image is presented in order
to show a clear picture of the jet as well as the main body of the galaxy).
In the middle and bottom panels the 6.7 $\mu m$ and 15 $\mu m$ maps (6'' pixel size)
are shown respectively, along with
contours of the same maps overlaid to emphasize the shape of the MIR images.
The contour levels were chosen to provide a good idea of the flux distribution
in the fainter extended regions as well as the central regions where the 
flux peaks.
The contour levels for each galaxy are given in the figure captions of
Figs. A.1 - A.9.

\subsection{15 $\mu m$ emission maps and dust seen in extinction}

It is possible that the MIR emission of early-type galaxies is associated with
continuum emission of very small grains (VSGs) which, if present, are heated to temperatures of a few 
hundred degrees (Madden et al. 1999, Ferrari et al. 2002). VSGs together with other
types of dust grains, e.g. PAHs and bigger grains (with sizes from 15 nm to 110 nm; D\'{e}sert et al. 1990)
which peak at FIR wavelengths,
should also contribute
to the extinction of the optical light and thus show up
as extinction features. 
Comparing optical and MIR images of spiral galaxies has been proven to 
be a very useful tool for mapping the distribution of dust grains of
all sizes throughout the images of the galaxies (Block et al., 1997; Block \& 
Sauvage 2000).
A major difficulty in comparing extinction features and 
the MIR emission in early-type galaxies is the uncertainty of the way that
dust is distributed inside those galaxies. As an example one can think of the dust
distributed in a smooth and homogeneous way relative to the stars and thus  being invisible 
when looking at the extinction of the starlight. In some galaxies though, HST has revealed
extinction features in the cores of early-type galaxies (van Dokkum \& Franx 1995). 

ISO is still the most sensitive instrument 
to date for MIR mapping. In Fig. 2, we compare the MIR emission at 15 $\mu m$ 
with extinction features that we see in the optical. The background image
is an HST observation of the core of the galaxy Fornax-A (NGC 1316).
We compare this image with
the MIR emission from this galaxy by overplotting the 15 $\mu m$ contours.
We see that the 15 $\mu m$ emission is clearly 
associated with the dust extinction features.
The outer right part of the image shows a large arc-like
structure (region A) traced by the 15 $\mu m$ emission. In regions
B and C, optical dust patches are shown to be well correlated with the shape of the
overlaid 15 $\mu m$ emission contours. A much more obvious correlation 
between
the 15 $\mu m$ emission and the dust as seen in extinction is shown in the
galaxy NGC 5266 (Fig. 3) where large scale extinction
lanes indicate the presence of a dust disk. This dust disk is very
obvious in the 15 $\mu m$ contours.

\subsection{SEDs}
In order to broaden our understanding of the nature and the properties of the ISM
in early-type galaxies it is important to properly model the Interstellar Radiation 
Field (ISRF) and thus to infer any excess that may be present in the MIR.
We do that by modelling the SED of all the galaxies in this sample using all the
available data from the UV to the FIR and the stellar evolutionary model P\'EGASE
(Fioc \& Rocca-Volmerange 1997).

\subsubsection{The data}
Optical and NIR data were obtained from databases where aperture photometry has been reported.
For all the galaxies (except for NGC 185 and NGC 205) the optical data (in most
cases a full set of U, B, V, R and I aperture photometry)
were taken from
Prugniel \& Heraudeau (1998).
For NGC 185, B, V and R aperture photometry were taken from Sandage \& Visvanathan (1978)
while for NGC 205, V, J, H and K aperture photometry were obtained from Frogel et al. (1978).
Near-infrared J, H and K aperture photometry were then taken
from Persson et al. (1979) for the galaxies NGC 1316, NGC 1399, NGC 4374, NGC 4473,
NGC 4486, NGC 5018, NGC 5102, NGC 5363 and from Frogel et al. (1978) for NGC 2300,
NGC 4278 and NGC 4649.
For the galaxies NGC 1052, NGC 3928, NGC 5173 and NGC 5866 we obtained J, H and K aperture
photometry from the 2MASS database.
Both the optical 
and NIR observations were then corrected for foreground galactic extinction by applying a
galactic extinction law with values of the B-band Galactic extinction taken from
the RC3 catalogue (de Vaucouleurs et al. 1991).
The FIR data at 60 and 100 $\mu m$ was taken from the IRAS catalogue.
The NED and SIMBAD databases were used extensively to obtain the data.

For the MIR data, presented in this work, the integrated flux was measured
for all the galaxies and all the filters in concentric apertures centered
at the maximum flux (the galaxy center) with an increasing
radius of one pixel. In this way, a growth curve of the galaxy's flux was
constructed out to distances limited by the frame size.
The photometric errors were calculated by taking into account the errors
introduced by the photon noise, the readout noise, the memory effects 
as well as a 5\% error due to flux density calibration uncertainties (ISOCAM handbook) and an
additional 5\% variation along each orbit (ISOCAM handbook). A more detailed
description of the evaluation of the photometric errors is given in 
Roussel et al. (2001). The total fluxes (integrated within the field of view
of ISOCAM) for each galaxy are given in Table 2.

%%%%%%%%%%%%%% TAB 2
\begin{table}
\begin{center}
\caption{Total integrated MIR (4.5 $\mu m$, 6.7 $\mu m$ and 15 $\mu m$) 
fluxes inside the field of view for the
galaxies in the sample.}
\begin{tabular}{lccc}\hline \hline
NAME & 4.5 $\mu m$ & 6.7 $\mu m$ & 15 $\mu m$ \\
     & (mJy)       & (mJy)        & (mJy)      \\
\hline
NGC 185   & 246.5 $\pm$ 20.4  & 76.7   $\pm$ 6.3  & 59.3   $\pm$ 6.7   \\
NGC 205   & 234.1 $\pm$ 19.7  & 111.5  $\pm$ 8.9  & 23.5   $\pm$ 5.1   \\
NGC 1052  & 249.8 $\pm$ 29.6  & 140.2  $\pm$ 13.9 & 230.6  $\pm$ 63.9  \\
NGC 1316  & 776.2 $\pm$ 57.6  & 410.4  $\pm$ 31.1 & 210.1  $\pm$ 34.0  \\
NGC 1399  & 414.7 $\pm$ 35.4  & 178.7  $\pm$ 14.4 & 88.6   $\pm$ 25.4  \\
NGC 2300  & 149.2 $\pm$ 21.8  & 80.8   $\pm$ 8.8  & 26.5   $\pm$ 17.4  \\
NGC 3928  & 42.5  $\pm$ 18.0  & 154.6  $\pm$ 14.4 & 152.2  $\pm$ 26.5  \\
NGC 4278  & 267.7 $\pm$ 26.8  & 144.7  $\pm$ 12.4 & 45.0   $\pm$ 13.8  \\
NGC 4374  & 546.5 $\pm$ 43.3  & 219.4  $\pm$ 18.6 & 89.6   $\pm$ 23.6  \\
NGC 4473  & 332.7 $\pm$ 32.4  & 127.1  $\pm$ 15.6 & 74.8   $\pm$ 29.4  \\
NGC 4486  & 759.3 $\pm$ 56.1  & 440.9  $\pm$ 32.1 & 105.9  $\pm$ 21.2  \\
NGC 4649  & 691.6 $\pm$ 50.1  & 308.1  $\pm$ 24.1 & 58.4   $\pm$ 10.8  \\
NGC 5018  & 242.6 $\pm$ 25.7  & 123.6  $\pm$ 13.5 & 54.3   $\pm$ 16.3  \\
NGC 5102  & 244.9 $\pm$ 22.0  & 109.2  $\pm$ 10.9 & 30.8   $\pm$ 23.1  \\
NGC 5173  & 26.0  $\pm$ 13.3  & 29.7   $\pm$ 7.5  & 23.9   $\pm$ 14.4  \\
NGC 5266  & 197.6 $\pm$ 21.9  & 149.8  $\pm$ 12.8 & 115.4  $\pm$ 28.5  \\
NGC 5363  & 384.4 $\pm$ 29.3  & 311.0  $\pm$ 23.4 & 99.8   $\pm$ 17.2  \\
NGC 5866  & 358.4 $\pm$ 31.6  & 262.1  $\pm$ 20.6& 179.6  $\pm$ 23.5  \\
\hline
\end{tabular}
\end{center}
\end{table}

After having collected all the data described above, a common aperture was chosen for each galaxy
(mainly introduced from the NIR data which, in many cases,
were limited to a few aperture sizes). This common aperture is 60'' (NGC 185), 48'' (NGC 205), 
75'' (NGC 1052), 56'' (NGC 1316), 
56'' (NGC 1399), 32'' (NGC 2300), 42'' (NGC 3928), 32'' (NGC 4278), 56'' (NGC 4374), 
56'' (NGC 4473), 56'' (NGC 4486), 48'' (NGC 4649), 56'' (NGC 5018), 56'' (NGC 5102), 
22'' (NGC 5173), 87'' (NGC 5266), 56'' (NGC 5363) and 99''(NGC 5866).
In order to
find the flux values corresponding to this aperture an interpolation routine was
applied to the flux growth curve for each filter. This was not possible 
for the IRAS-FIR data where only one flux measurement for the total galaxy was available.
The aperture fluxes for each galaxy are shown in Fig. 4 (points) together with
the modelling that was applied to these points (see below).
%
%%%%%%%%%%%%%% TAB 3
\begin{table}
\begin{center}
\caption{Model parameters. The age and metallicity as calculated from
the P\'EGASE stellar evolutionary model, and the dust mass and dust temperature
as derived from black-body fits, assuming an emissivity index of $\beta = 1$.}
\begin{tabular}{lcccc}\hline \hline
NAME & Metallicity & Age & Dust Temp. & Dust mass  \\
     &             &(Myr)&  (K)       & M$_{\sun}$ \\
\hline
NGC 185   & 0.01   & 1400   & 26  & 865   \\
NGC 205   & 0.02   & 350    & 24  & 3.4  $\times 10^3$ \\
NGC 1052  & 0.10   & 7000   & 35  & 1.4  $\times 10^5$ \\
NGC 1316  & 0.03   & 4000   & 30  & 1.5  $\times 10^6$ \\
NGC 1399  & 0.04   & 4000   & 21  & 4.3  $\times 10^5$ \\
NGC 2300  & 0.02   & 14000  & 28  & 5.5  $\times 10^4$ \\
NGC 3928  & 0.04   & 1600   & 35  & 4.6  $\times 10^5$ \\
NGC 4278  & 0.02   & 8000   & 30  & 2.7  $\times 10^5$ \\
NGC 4374  & 0.03   & 4500   & 33  & 1.0  $\times 10^5$ \\
NGC 4473  & 0.02   & 8000   & 36  & 6.8  $\times 10^3$ \\
NGC 4486  & 0.04   & 3000   & 45  & 1.1  $\times 10^4$ \\
NGC 4649  & 0.04   & 3000   & 40  & 4.5  $\times 10^4$ \\
NGC 5018  & 0.02   & 8000   & 35  & 4.7  $\times 10^5$ \\
NGC 5102  & 0.03   & 1000   & 30  & 2.6  $\times 10^4$ \\
NGC 5173  & 0.008  & 9000   & 37  & 1.4  $\times 10^5$ \\
NGC 5266  & 0.03   & 3000   & 28  & 1.1  $\times 10^7$ \\
NGC 5363  & 0.02   & 10000  & 30  & 6.9  $\times 10^5$ \\
NGC 5866  & 0.05   & 2000   & 28  & 2.5  $\times 10^6$ \\
\hline
\end{tabular}
\end{center}
\end{table}

\subsubsection{The model}
The optical light of early-type galaxies originates primarily from evolved stars.
Fits to the
optical and NIR data points have shown that the stellar component of these galaxies
can be approximated by simple black bodies with temperatures ranging
from T$\sim 3000$ K to T$\sim 5000$ K (Ferrari et al. 2002, Athey et al. 2002).
In order to have a more detailed description of the stellar and dust
content of early-type galaxies,
we model the SEDs from the UV to the MIR
using the P\'EGASE
evolutionary synthesis model (Fioc \& Rocca-Volmerange 1997). Given
a certain age for the stellar population, an initial mass function (IMF)
and a metallicity,
P\'EGASE computes the stellar SED. For the formation of the stars
we assume an instantaneous burst event. Extinction of the stellar
light by the dust is not taken into account since we assume that the amount
of dust in these galaxies is insufficient to produce any
serious obscuration effects. We caution the reader however that in some
cases (e.g. NGC 5266 and NGC 5866) where the dust is seen in absorption,
forming large scale disks, the extinction effects may be significant
and thus proper treatment using a radiative transfer model is needed.
In order to model the stellar SED
that best fits our data, we minimize the $\chi^2$ allowing for 
a variation of the initial metallicity and the age of the galaxy.
Although this method gives uncertainties in the ages and the metallicities 
of the galaxies
(compared to a more direct determination 
of these parameters from the actual spectra of the galaxies, e.g.
Caldwell et al. 2003), it does provide an indication of the range
of the values of these parameters. The MIR domain of the stellar SED 
though ($\lambda \ge 2 \mu m$), where we
investigate the MIR excess, is somewhat insensitive to these
parameters.
In each galaxy, the data set that was considered to be representative
of the stellar contribution to the SED contained all of the available 
observations ranging from UV to 4.5 $\mu m$.
The FIR observations were fitted with a modified black-body
with the emissivity index $\beta = 1$ (which is the value that is often used
for $\lambda < 250 \mu m$; see Alton et al. 1998) allowing for the determination
of dust temperatures as well as dust masses.
The SEDs are shown in Fig. 4. The points represent the fluxes in a given aperture
while the models for the stellar radiation field (left curve) and for the 
emission of the dust in the FIR (right curve) as well as the total SED 
(on top of the other components) are shown
with solid lines. Also the relative difference between observation and model is given
in the bottom panel for each galaxy.
In Table 3 the model parameters obtained from P\'EGASE (age and metallicity) and by
modelling the FIR (dust temperatures and dust masses) are reported. 

%%%%%%%%%%%%%% TAB 4
\begin{table}
\begin{center}
\caption{The MIR excess at $6.7 \mu m$ and $15 \mu m$ for each galaxy.
The excess is quantified as the excess beyond the stellar component which is modelled using the stellar
evolutionary synthesis model P\'EGASE.}
\begin{tabular}{lcc}\hline \hline
NAME & $6.7 \mu m$ excess & $15 \mu m$ excess  \\
     &      (mJy)          &      (mJy)         \\
\hline
NGC 185   & 9.20$ \pm$ 2.94   & 23.65 $ \pm$ 2.64   \\
NGC 205   & 8.95 $ \pm$ 3.59 &  6.36  $ \pm$ 5.9 \\
NGC 1052  & 12.38$ \pm$ 17.86  & 181.81  $ \pm$ 76.70 \\
NGC 1316  & 38.17 $ \pm$ 22.95 & 97.42 $ \pm$  33.96 \\
NGC 1399  & -5.79 $ \pm$ 11.43 &  23.34  $ \pm$ 29.00\\
NGC 2300  & -0.90 $ \pm$ 2.89 & $ 5.13\pm$ 5.08 \\
NGC 3928  & 47.55 $ \pm$ 7.02 &56.38  $ \pm$ 10.13 \\
NGC 4278  & 4.77 $ \pm$ 5.32 & 19.42 $ \pm$4.79 \\
NGC 4374  & -2.24$ \pm$ 12.48&  28.88  $ \pm$ 9.08\\
NGC 4473  & -4.32 $ \pm$ 4.96 & 11.05  $ \pm$ 5.52\\
NGC 4486  &  35.48 $ \pm$ 15.89& 51.32  $ \pm$ 11.82 \\
NGC 4649  & 5.14 $ \pm$ 13.41 &  5.44 $ \pm$ 6.14 \\
NGC 5018  & 11.25 $ \pm$ 6.49 & 24.96 $ \pm$ 5.76  \\
NGC 5102  & 4.96$ \pm$ 6.33  & 14.75  $ \pm$ 14.03 \\
NGC 5173  & 7.23$ \pm$ 1.38 & 13.29  $ \pm$ 2.42 \\
NGC 5266  & 40.72 $ \pm$7.20  & 69.51 $ \pm$ 8.65  \\
NGC 5363  & 55.47 $ \pm$11.99  & 68.73 $ \pm$ 10.21 \\
NGC 5866  & 87.62 $ \pm$ 19.35 &139.65 $ \pm$ 21.01  \\
\hline
\end{tabular}
\end{center}
\end{table}

%%%%%%%%%%%%%%%%%%%%%%%%%%%%%%%%%%%
\section{Discussion}
%%%%%%%%%%%%%%%%%%%%%%%%%%%%%%%%%%%
\subsection{IR emission}
The IR SED traces various compositions and size distributions of dust and also reflects
the environments in which dust is embedded. In the FIR wavelength regime,
dust grains with typical sizes thought to range from 0.015 to 0.110 $\mu m$ in
our Galaxy (D\'{e}sert et al.
1990) are heated to a range of temperatures approximated here by an `average'
black-body modified by an emissivity index $\beta = 1$ 
with temperature ranging from $\sim 20$ to 40 K.
The FIR SED is limited in wavelength coverage only to the IRAS bands (except
for NGC 205 where ISO observations in the FIR from Haas (1998) are also presented)
and thus could be missing colder dust traced by submillimeter and millimeter
observations
(Wiklind \& Henkel 1995). Thus, our derived dust masses (listed in Table 3) are lower
limits to the dust masses.
Sampling these galaxies at submillimeter wavelenghts would give us a more 
accurate picture of any colder dust that might be present (i.e. Klaas et al. 2001;
Galliano et al. 2003a, Galliano et al. 2003b).

The MIR part of the SED on the other hand is a more complex regime (see Sect. 4)
where a smooth
turnover from stellar emission to emission from thermally fluctuating PAHs
and very
small dust grains takes place (see D\'{e}sert et al.
1990). 

After accounting for the stellar contribution in the MIR we find a clear 
excess of emission in the 15 $\mu m$ band in most of our galaxies.
The MIR excess is plotted in relative values in the panel below the 
individual SEDs of Fig. 4.
We also report the absolute 6.7 and 15 $\mu m$ excess for each galaxy in Table 4.
The excess is given in mJy and the associated error is the photometric error
for each measurement. This means for example that the 6.7 $\mu m$ excess 
over the stellar emission for NGC 185 is between 6.26 mJy and 12.14 
mJy if we take into
account the photometric errors. 
In NGC 1399 and NGC 5102 where a much more complete set of data
is available (Fig. 4) we see an emission excess from 9.6 - 15 $\mu m$ which
coincides with the 9.7 $\mu m$ silicate dust band (see also Athey et al., 2002).
If we now consider only those measurements 
that show a clear excess we conclude that out of the 18 galaxies in this
sample, 10 show excess at  6.7 $\mu m$ and thus possible emission from
PAH features and 14 show excess at 15 $\mu m$, presumably from small hot grains.

\subsection{morphology}
Examining the  morphology of the MIR maps provide us with significant
information on the origin of the dust. If dust has an internal origin
(i.e. coming from mass loss from asymptotic giant branch (AGB) stars), then
the spatial distribution of the dust should resemble that of the
stars. The purely stellar distribution (traced by the 4.5 $\mu m$
observations) in early-type galaxies generally
follows a de Vaucouleurs law (de Vaucouleurs 1953). Any deviation from
this behavior of the MIR emission should in principle, indicate another anomalous distribution 
of dust (e.g. disks, dust lanes, dust patches, etc.) and thus
an external origin  of the dust (i.e. due to merging and interaction 
events). We investigate this by plotting the MIR (4.5 $\mu m$,
6.7 $\mu m$, 15 $\mu m$) azimuthally
averaged radial profiles of the galaxies,
using the 4.5 $\mu m$ distribution
as a reference of the purely stellar content of the galaxy. 
We do this by using the {\it fit/ell3} task of MIDAS which is based
on the formulas of Bender \& M\"{o}llenhof (1987).
This routine computes the elliptical isophotes of the
galaxy and averages the flux over the ellipses. Then, the averaged
flux is given as a function of the distance to the major axis
of the galaxy. This is presented in Fig. 5 with the radial profiles
for each galaxy and for the three bands: 4.5 $\mu m$ (circles)
6.7 $\mu m$ (boxes) and 15 $\mu m$ (stars) plotted against
the distance to the major axis in the power of 1/4 (in kpc). The fluxes are given
in mJy but for reasons of clarity, a few profiles have been shifted
(see figure caption of Fig. 5). In this
configuration, a de Vaucouleurs profile is
a straight line. 
In Fig. 6 we present the azimuthally averaged MIR flux ratio profiles.
The $6.7 \mu m / 4.5 \mu m$ ratio is plotted with boxes while the
$15 \mu m/ 4.5 \mu m$ ratio is plotted with stars for each galaxy.
We normalize the MIR flux to the stellar flux
(in this case the 4.5 $\mu m$ flux) to trace the ISM.
For example if the $15 \mu m/ 4.5 \mu m$ ratio increases
in the center compared to the outskirts of the galaxy
this would indicate that the dust is concentrated toward the center.
Furthermore, these MIR flux ratios are good indicators of the stellar
contribution in these galaxies. In regions where galaxies are totally
devoid of ISM emitting in the MIR (and thus showing no excess 
over the stellar component), these ratios should take values characteristic
of old stellar populations. In each of the plots of Fig. 6 we indicate
the values of these ratios for a black-body of 3000 K for example. The $6.7 \mu m / 4.5 \mu m$ ratio
is 0.55, for a black-body of 3000 K and
is indicated by a solid arrow while the $15 \mu m/ 4.5 \mu m$ ratio is 0.14 and
is indicated
by an open arrow.

From these plots we see that:

\begin{itemize}
\item  All of the 4.5 $\mu m$ observations,
except for NGC 4486 (M87)
and NGC 5866,
can be well described with a de Vaucouleurs profile.
This is also evident in the 4.5 $\mu m$ maps overlaid on the optical image
of the galaxy (see the figures in Appendix A) where in all of the cases, 
there is a very good correlation between the morphology of the optical
image and the 4.5 $\mu m$ emission contours. In the case of
NGC 4486 the jet is very prominent in the 4.5 $\mu m$ image, while NGC 5866 shows an edge-on
disk. 

\item The proximity of the two nearby dwarf elliptical galaxies, NGC 185 and NGC 205, 
gives a pc-scale resolution which reveals
a clumpy distribution of the
6.7 and 15 $\mu m$ emission. This can be seen in both the radial profiles
(Fig. 5) and the flux ratio profiles (Fig. 6) 
as well as in the individual maps shown in Fig. A.1. The emission is
associated with individual dust patches seen in extinction 
(Welch et al. 1998, Martinez-Delgado et al. 1999).

\item NGC 4486 (M87) shows MIR emission which is
associated with the jet in all MIR bands (see also the maps in Fig. A.6).
As we shall see later (Sect. 6.4) the MIR excess is dominated by the 
synchrotron emission of the jet.

\item NGC 1052 shows centrally concentrated emission at 6.7 and 15 $\mu m$
due to a compact central source (believed to be an AGN). This becomes more evident in
Fig. 6 where there is a large excess of the flux ratios close to the 
center.

\item NGC 5266 and NGC 5866 show large scale disks. This is shown in Fig. 5
with the radial profiles deviating from the de Vaucouleurs profile
and in Fig. 6 with the MIR flux ratio profiles showing
peculiar distributions.

\item NGC 1316, NGC 3928 and NGC 5363 show MIR flux ratio profiles 
that take on higher values close to the center of the galaxy (Fig. 6)
indicating that the ISM is more concentrated 
toward the center.

\item The other galaxies show radial profiles for the 6.7 and 15 $\mu m$
observations which are close to the de Vaucouleurs profile (Fig. 5). 
They also show more or less flat MIR flux ratio profiles (Fig. 6)
indicating that in those galaxies where dust exists
the distribution follows the stellar distribution.

\item Most of the galaxies have asymptotic values at large radii for the flux ratios 
which are close to the characteristic of old stellar population.
\end{itemize}

Summarizing all the above considerations, we can conclude that out of 18 galaxies
studied here, 2 dwarf ellipticals show a clumpy distribution of dust,
2 galaxies show large scale disks, 2 galaxies are devoid of dust,
1 shows compact MIR emission due to an AGN, 1 shows emission 
associated with a jet, 3 show MIR emission which is more 
centrally concentrated and the remaining 7 galaxies show smooth
distributions of the dust throughout the galaxy.
Considering the small number of galaxies and the variety of types of galaxies
in this sample we definitely can not draw any general statistical conclusion 
about the dust distribution.
We show, however, many different morphologies that an early-type galaxy may
exhibit in the MIR wavelengths supporting all different scenarios of the 
origin of the dust (already mentioned in the introduction).

\subsection{Colour-Colour diagrams}

The ratio of the total 6.7 $\mu m$ flux over the 15 $\mu m$ flux has been used
to interpret the activity of the ISM in galaxies (Madden et al. 1999,
Dale et al. 2000, Vaisanen et al. 2002). In these studies a large sample
of galaxies with only a small sub-sample of early-type galaxies is
used to delineate regions of different groups of galaxies and thus
infer diagnostics for the behavior of the ISM. 
We repeat this kind of diagnostic color-color diagrams but now
with more data of early-type galaxies in order to investigate 
their properties in terms of their infrared colors and how they differ
from other type of galaxies.
In Fig. 7 we compare the
$F_\nu(6.7 \mu m)/F_\nu(15 \mu m)$ ratio with the $F_\nu(12 \mu m)/F_\nu(25 \mu m)$
ratio for three samples of galaxies. The first sample (solid circles) is a sample
of normal star-forming galaxies (Dale et al. 2000), the second sample
(open squares) is another (complementary to the first one) sample of
spiral galaxies (Roussel et al. 2001) while the third sample (red triangles)
consists of the sample of early-type galaxies presented in this work.
A cross on the top left part of the diagram shows a typical error bar
associated with the early-type galaxies.
As we can see there is a clear trend with the type of the galaxy becoming
earlier when going upwards along the diagonal. This is to be expected and
it is indicative of the dearth of ISM relative to the importance
of the stellar contribution in the MIR in early-type systems
compared to late-type ones.
Late-type galaxies show
larger values of $F_\nu(15 \mu m)$ and $F_\nu(25 \mu m)$ compared
to the $F_\nu(6.7 \mu m)$ and $F_\nu(12 \mu m)$ respectively, due to
larger contribution from higher temperature dust as a result of
more active star formation.

In Fig. 8 we present the same samples (with the same symbol coding as in Fig. 7)
but now the MIR color $F_\nu(6.7 \mu m)/F_\nu(15 \mu m)$
is plotted with the FIR color $F_\nu(60 \mu m)/F_\nu(100 \mu m)$
for each galaxy. 
The cross on the top right part of the diagram shows a typical error bar
associated with the early-type galaxies.
Here, it is interesting to notice the clear trend for
the spiral galaxies (already discussed in Dale et al. 2000) with the MIR
color being roughly constant and near unity going from quiescent to midly
active galaxies (close to the point where the FIR color reaches
the value of -0.2) and then the drop of the MIR color when
going to more active galaxies (with FIR color larger than -0.2).
It is interesting to notice in this diagram that, except for the
dustier early-type galaxies all the rest of the early-type galaxies 
lie in an area which is above 
that occupied by the spirals, due to the diminishing 15 $\mu m$ flux,
which is again indicative of the lack of
the ISM in these galaxies with respect to the disk galaxies. 

\subsection{Synchrotron emission in the IR}

The exact knowledge of the quantity and the characteristics of the MIR excess is an
important diagnostic to address the existence, nature and composition of the dust. 
Complications arise when examining active galaxies where
the MIR excess could be attributed to several components such as the galactic dust,
emission from stellar photospheres, thermal emission from the nuclear torus and
synchrotron emission from jets that are hosted inside the galaxies. In this section
we examine the possibility that synchrotron emission may contribute to the 
MIR flux by
modelling the SED of the well known galaxy NGC 4486 (M87) which hosts 
a very powerful jet in its center (see Sect. 6.5 for more details).

We have gathered all the available radio observations (mainly using the
NED database) and plotted together with all the other observations (UV, optical,
NIR, MIR, FIR)
as described in Sect. 5.3.1 (see Fig. 9).
We then used a simple power-law in order
to account for the radio emission and we extrapolate from the radio to the MIR regime. 
This extrapolation is valid from a theoretical point of view and it has been
used in order to explain the emission of jets in a wide range of wavelengths
(Perlman et al. 2001, Cheung 2002, Marshall et al. 2002).
In Fig. 9 we show the stellar radiation field (as computed by P\'EGASE; see 
Sect. 5.3.2), the modified black-body law which models the emission of the warm dust
detected by IRAS, the radio emission using a simple power law, namely
$\nu$F$_\nu \propto \nu^{\alpha}$, as well as the total predicted SED which is 
the sum of all of the 
components of the SED. 
For the power-law component we derive a value of $\alpha = 0.16$ which
is indicative of synchrotron emission (Condon 1992) while a temperature of 58 K is found
for the modified black-body peaking in the FIR in order to fit the FIR data.

It is obvious that in this galaxy the MIR excess could easily be
explained by the contribution of the synchrotron radiation. This is also nicely
presented in the lower panel where the relative difference between
the total SED model and the observations is given. 
We can see now that this three component model (evolved stellar population + dust emission
described by a modified black-body + radio emission described by a power law) fits all of the data,
including the MIR, in contrast to that presented in Fig. 4.
With this example we want to emphasize the fact that synchrotron radiation
may play a very significant role even in the IR spectrum of strong radio 
galaxies.

\subsection{Notes on individual galaxies}

\hspace{0.5 truecm}{\bf NGC 185.}
This dwarf elliptical galaxy is of great interest not only because of its
proximity to us (only $\sim 0.7$ Mpc away) but also because it contains 
optical dust clouds (Martinez-Delgado et al., 1999) and luminous blue stars
(Baade 1944, Lee et al. 1993) indicating current star formation activity.
This galaxy shows a centrally concentrated  distribution for the atomic hydrogen which
extends further to the northeast while CO observations show that the molecular
gas is closely associated with the optical dust clouds (Welch et al. 1996, Young \& Lo
1997, Young 2001). The $15 \mu m$ map presented in this work (Fig. A.1) is mainly
associated with the dust clouds and the molecular gas but also shows a central 
concentration.

{\bf NGC 205.}
This is also a dwarf elliptical galaxy which shows similar characteristics to
NGC 185. The bright blue stars observed in this galaxy (Hodge 1973) as well as
detailed study of the color profiles (Peletier 1993), show that star formation
activity is still taking place in this galaxy.
The atomic hydrogen in this galaxy is distributed in an elongated
structure which is curved and extends about twice as far to the south of the optical center
as it does to the north of the optical center (Young \& Lo 1997). CO observations 
show that the molecular gas is associated with the dust clouds (Welch et al. 1998,
Young \& Lo 1996, Young \& Lo 1997). The $15 \mu m$ emission (Fig. A.1) is associated
with the dust clouds.

{\bf NGC 1052.}
This giant elliptical galaxy which is classified as a low-ionization nuclear emission-line
region (LINER) galaxy (Ho et al. 1997) hosts a double-sided radio jet with a size of
a few milliarcseconds (Kameno et al. 2002). HST observations 
have revealed dust absorption features in the center of this galaxy (van Dokkum et al. 1995). 
In the MIR maps of Fig. A.2 it is clearly seen that 
the $15 \mu m$ emission is concentrated in a very compact region
close to the center and is dominated by the central source which, in this case, is an
active nucleus in agreement with the findings of
Knapp et al. (1992). This is also seen in the $6.7$ and $15 \mu m$ radial profiles (Fig. 5) 
as well as in the MIR flux ratio profiles (Fig. 6)
where a steep turnover is present at a distance close to the nucleus.

{\bf NGC 1316.}
This is a giant elliptical galaxy in the Fornax cluster showing evidence 
of recent merging events (e.g. dust patches,
$H\alpha$ filaments, ripples and loops; 
see, e.g. Schweizer 1980).
Observations of atomic and molecular gas indicate an absence of neutral atomic gas in the
galaxy (except for four bright HII regions) while there was a significant amount 
of molecular gas ($\sim 5 \times 10^8 M\sun$) which is mainly associated with the dust
patches which are oriented along the minor axis (Horellou et al. 2001).
ROSAT observations have shown that the X-ray emission is extended and flattened along the major
axis of the galaxy which indicates the presence of a hot gas component (Kim et al. 1998). 
The 15 $\mu m$ emission is oriented along the minor axis of the galaxy and traces
the dust patches quite well in the outer parts (see Fig. 2).

{\bf NGC 1399.}
This is the central dominant galaxy of the Fornax cluster which has been extensively studied
in a wide range of wavelengths. It hosts a low-luminosity radio jet which is confined 
within the optical image (Killeen et al. 1988) and it shows an extended and asymmetric
gaseous halo in X-rays (Paolillo et al. 2002). Although there is no detection of
dust seen by optical obscuration (van Dokkum \& Franx 1995), we observe
a MIR excess between the 9.62 and 15 $\mu m$ bands
(Fig. 4). This excess (as already reported in Athey et al. 2002)
coincides with the 9.7 $\mu m$ silicate dust band that originates in the 
circumstellar envelopes of AGB stars.

{\bf NGC 2300.}
HST observations show no signs of dust in this galaxy. 
Our analysis shows that there is no significant excess observed in the
MIR bands supporting the lack of ISM in this galaxy.

{\bf NGC 3928.}
NGC 3928 is an early type gas-rich starburst galaxy with CO detected in a
rotating disk. The MIR spectrum of this galaxy shows a striking excess over
the stellar emission, suggesting a large content of warm dust
and PAHs.

{\bf NGC 4278.}
This elliptical galaxy shows
a complex dust structure in its core
(van Dokkum \& Franx 1995). We do not detect any excess at 6.7 $\mu m$ but we
detect a 15 $\mu m$ excess indicating the presence of warm dust in this galaxy.

{\bf NGC 4374.}
NGC 4374 is a giant elliptical galaxy which show strong radio emission and a two-sided
jet emerging from its compact core. HST images reveal dust lanes in the core of the galaxy
while no significant amount of diffusely distributed cold dust was detected at submillimeter
wavelengths (Leeuw et al. 2000). 
As in NGC 4278 we do not detect any excess at 6.7 $\mu m$ but we
detect 15 $\mu m$ excess.

{\bf NGC 4473.}
Although there is no detection of dust seen in optical absorption for this galaxy
(van Dokkum \& Franx 1995) we detect extended MIR emission at 15 $\mu m$ which could be
associated with a smooth dust component, preventing detection in the optical.

{\bf NGC 4486.}
This giant elliptical galaxy which is located at the center of the Virgo cluster,
hosts a very strong central radio source and a synchrotron jet which is visible
from radio to X-ray wavelengths (Marshall et al. 2002). The nucleus of this galaxy
show no direct signatures of dust (van Dokkum \& Franx 1995)
while both molecular and atomic gas components are absent from the center of this 
galaxy (Braine \& Wiklind 1993). 
The MIR maps of this galaxy presented in Fig. A.6, show a central emission
component as well as a separate component that traces the jet morphology.
We have shown that 
all the MIR excess and most
of the FIR output of this galaxy as a whole (galaxy + radio jet) can be 
attributed to a single synchrotron emission component as seen in Fig. 9.

{\bf NGC 4649.}
NGC 4649 shows no clear excess in the MIR (see Fig. 4 and Table 4).
IRAS fluxes indicate a dust mass of $4.5 \times 10^4 M\sun$ (see Table
3) although, as already stated in the introduction, background contamination
and other uncertainties may lead to an overestimation of the FIR output 
of early-type galaxies (Bregman et al. 1998).
The absence of any
significant amount of dust is also demonstrated by recent ISO observations
at 60, 90 and 180 $\mu m$ where this galaxy show no emission (Temi et al. 2003).

{\bf NGC 5018.}
This is a luminous metal-poor giant elliptical (Bertola et al. 1993). HST images
reveal complex but very localized dust structures.  
With our analysis we detect both 6.7 and 15 $\mu m$ excesses from this galaxy.

{\bf NGC 5102.}
NGC 5102 is a gas-rich lenticular galaxy that shows evidence for recent 
star formation activity in the central region (Danks et al. 1979, Deharveng et al.
1997). HST observations show dust signatures in the core of this galaxy
(van Dokkum \& Franx 1995).
In the SED of this galaxy (with a wide range of observations available) presented in 
Fig. 4 we see a clear excess over
the stellar emission between the 9.62 and the 15 $\mu m$ bands. 
As in the case of NGC 1399 this coincides with the 
9.7 $\mu m$ silicate dust band (see also Athey et al. 2002).

{\bf NGC 5173.}
This galaxy shows an extended emission in the MIR distributed in a smooth
way and peaked at the center. HST observations show a high filamentary structure 
in the core of this galaxy (van Dokkum \& Franx 1995). From its SED (Fig. 4)
we see prominent dust excess over the stellar emission.

{\bf NGC 5266.}
NGC 5266 is an HI-rich elliptical galaxy with a dust ring along the minor
axis of the galaxy (Varnas et al. 1987, Morganti et al. 1997). From the 
MIR maps of Fig. 3 and Fig. A.8 we can see a disk at 15 $\mu m$ (slightly warped) which
is aligned with the minor axis of the galaxy.
The SED shows a clear 
excess of the MIR emission over the 
stellar component.

{\bf NGC 5363.}
This elliptical galaxy shows a dust lane along its minor axis. This is also
shown in the MIR maps (Fig. A.9) with the 15 $\mu m$ emission being aligned
with the minor axis. The MIR SED
shows a clear excess over the stellar emission. 

{\bf NGC 5866.}
This edge-on lenticular galaxy shows a remarkable dust disk which is oriented 
exactly along the major axis of the galaxy. This is also shown with 
the MIR observations of Fig. A.9. The MIR SED shows a clear excess in
both the 6.7 and 15 $\mu m$ bands.
The PHT-S spectrum of this galaxy shows a broad emission feature 
that peaks at 7.9 $\mu m$ (Lu et al. 2003).
In Fig. 10 we present the MIR 
profiles along the major-axis for both the 6.7 $\mu m$ (dashed line) and the
15 $\mu m$ (solid line) emission. The profiles are very similar at both wavelengths
with peak emission at the nucleus and secondary maxima at $\simeq 4$ kpc from either 
side of the center. This symmetrical secondary peak may be an indication of
a ring-like distribution of dust in this galaxy.

%%%%%%%%%%%%%%%%%%%%%%%%%%%%%%%%%%%%%%%%%
\section{Summary}
%%%%%%%%%%%%%%%%%%%%%%%%%%%%%%%%%%%%%%%%%
Eighteen early-type galaxies observed with ISOCAM at 4.5, 6.7 and 15 $\mu m$ were
analyzed in this study aiming to determine the characteristics of the ISM in these
galaxies. We have modelled the SED of these galaxies using the stellar evolutionary synthesis model,
P\'EGASE, to fit the stellar component and a modified black-body to fit the FIR 
part of the SED. The excess of the MIR emission (at 6.7 and 15 $\mu m$) over the 
stellar component is then quantified and it is found that out of 18 galaxies 
10 show excess at 6.7 $\mu m$ and 14 at 15 $\mu m$.
Thus the presence of PAHs and small dust grains is established in early-type 
galaxies. In two galaxies (NGC 1399 and NGC 5102) the excess at 9.7 $\mu m$,
due to silicate dust emission is also seen.

The morphology of the galaxies is examined by visual inspection of the emission
maps at the three bands (4.5, 6.7 and 15 $\mu m$) and by plotting the 
azimuthally averaged radial profiles as well as the MIR flux ratio
profiles. From this analysis we conclude that: 

\begin{enumerate}

\item The 4.5 $\mu m$ emission
follows quite well the de Vaucouleurs profile for all the galaxies (except
NGC 4486 where a jet is seen and NGC 5866 which has an edge-on configuration).

\item The two dwarf elliptical galaxies (NGC 185 and NGC 205)
show patchy distributions of the MIR
emission. 

\item  NGC 1052 shows a compact MIR emission due to an AGN.

\item Three galaxies (NGC 1316, NGC 3928 and NGC 5363) 
show MIR emission which is concentrated close to the
center of the galaxy. 

\item Two galaxies (NGC 2300 and NGC 4649) are totally devoid of dust.

\item The remaining ten galaxies show a smooth distribution of dust throughout the galaxy.

\item At the outskirts of the galaxies the profiles take values which are close 
to the values of an old stellar population.

\end{enumerate}

With ISOCAM-IRAS color-color diagnostic diagrams, we delineate the regions occupied by different
types of galaxies. Early-type galaxies, in general, occupy areas which favor the
dust poor galaxies while the spiral galaxies occupy the regions 
where the ISM-rich galaxies are located.

Finally, for one of the galaxies, the strong radio galaxy NGC 4486 (M87), 
we show that the MIR excess as well as most of its FIR output
is coming from synchrotron emission emerging from its powerful jet.

\begin{acknowledgements}
We thank Suvi Gezari and Pierre Chanial for their help in the earliest
stages of the data reduction. We are also greatful to Ren\'{e} Gastaud and Koryo Okumura
for their help in the data reduction and for useful discussions.
This research has made use of the NASA/IPAC Extragalactic Database (NED) which is operated
by the Jet Propulsion Laboratory, California Institute of Technology, under contract
with the National Aeronautics and Space Administration.
We have made extensive use of the SIMBAD database, operated at CDS, Strasbourg, France.
\end{acknowledgements}

%%%%%%%%%%%%%%%%%%%%%%%%%%%%%%%%%%%%%%%%%%%%%%%%%%%%%%%

%\appendix
%\section{Mid-IR maps}

\newpage

\noindent
{\bf FIGURE CAPTIONS}
\\\\
%%%%%%%%% FIG 1
\noindent
{\bf FIGURE 1}\\
Azimuthally averaged radial profile of NGC 1399 at $4.5 \mu m$ (points).
The solid line gives the best fit to the data while the dot-dashed line gives
the de Vaucouleurs profile for the galaxy and the dashed line gives the
constant value of the background determined from the fit. 
\\\\
%%%%%%%%%%%%%% FIG. 2
\noindent
{\bf FIGURE 2}\\
Optical HST image (background image) with the 6'' 15 $\mu m$
emission map (contours) overlaid for the central region of NGC 1316.
The contour levels are 0.3, 0.35, 0.4, 0.6, 0.7, 1.0, 1.2, 4.0, 6.0, 12.0 mJy/pixel.
Regions A, B and C show extinction features in the optical image that are traced
nicely by the 15 $\mu m$ contours.
\\\\
%%%%%%%%%%%%%% FIG. 3
\noindent
{\bf FIGURE 3}\\
Optical DSS image (background image) of NGC 5266 with the 6'' 15 $\mu m$
emission map (contours) overlaid.
\\\\
%%%%%%%%%%%%%% FIG. 4
\noindent
{\bf FIGURE 4}\\
SEDs of the early-type galaxies in the sample.
The points give the observations within a specific aperture while the lines
give the modelled stellar radiation field (the left profile), the FIR dust emission (the right profile)
and the total SED (overlaid as the sum of the components).
For each galaxy the relative difference between observation
and model is given in the bottom panel which is a means to quantify the MIR dust and/or PAH
excess between 6 and 18 $\mu m$.
\\\\
%%%%%%%%%%%%%%%%%%%%% FIG. 5 %%%%%%%%%%%%%%%%%%%%%%%%%%%%%%%%
\noindent
{\bf FIGURE 5}\\
Azimuthally averaged profiles of the galaxies in the sample.
The circles trace the 4.5 $\mu m$ flux, the rectangles the 6.7  $\mu m$
flux and the stars the 15  $\mu m$ flux.
The radial distance is measured along the major axis and it is
given in kpc to the power of 1/4. In this configuration,
a de Vaucouleurs profile would appear as a straight line.
The fluxes are given in units of
mJy. For clarity reasons the 6.7 $\mu m$ profiles of NGC 1052, NGC 5173, NGC 5266, NGC 5363,
NGC 5866 are shifted by -0.1, -0.4, -0.3, -0.2, -0.4 dex respectively while the
15 $\mu m$ profiles of NGC 185, NGC 1052, NGC 3928, NGC 5018, NGC 5173, NGC 5266, NGC 5363,
NGC 5866 are shifted by -0.4, -0.8, +0.5, -0.3, -0.9, -0.7, -0.4, -0.7 dex respectively.
\\\\
%%%%%%%%%%%%%%%%%%%%% FIG. 6 %%%%%%%%%%%%%%%%%%%%%%%%%%%%%%%%
\noindent
{\bf FIGURE 6}\\
Azimuthally averaged MIR flux ratio profiles. The $6.7 \mu m / 4.5 \mu m$ ratio
is plotted with boxes while the $15 \mu m / 4.5 \mu m$ ratio is plotted with stars. The
arrows on the right side of the plot indicate the value of the flux ratios of
$6.7 \mu m / 4.5 \mu m$ and $15 \mu m / 4.5 \mu m$ for a black-body of a temperature
of 3000 K for which the ratio $6.7 \mu m / 4.5 \mu m$ is 0.55 (indicated by the solid arrow)
while the ratio $15 \mu m / 4.5 \mu m$ is 0.14 (indicated by the open arrow).
The radial distance is given in kpc to the power of 1/4.
%%%%%%%%%%%%%%%%%%%%%%%%%%%%%%%%%%%%%%%%%%%%%%%%%%%%%%%%%%%%%%
\\\\
%%%%%%%%%%%%%%%%%%%% FIG. 7 %%%%%%%%%%%%%%%%%%%%%%%%%%%%%%
\noindent
{\bf FIGURE 7}\\
The $F_\nu(6.7 \mu m)/F_\nu(15 \mu m)$ ratio
versus the $F_\nu(12 \mu m)/F_\nu(25 \mu m)$  ratio for
the Dale et al. (2000) sample of normal star-forming galaxies (solid
circles), the Roussel et al. (2001) sample of spiral galaxies
(open squares) and the early-type sample of galaxies from this paper (red triangles).
The cross on the top left part of the diagram indicates a typical
error bar associated with the early-type galaxies analyzed in this work
\\\\
%%%%%%%%%%%%%%%%%%%% FIG. 8 %%%%%%%%%%%%%%%%%%%%%%%%%%%%%%
\noindent
{\bf FIGURE 8}\\
The $F_\nu(6.7 \mu m)/F_\nu(15 \mu m)$ ratio
versus the $F_\nu(60 \mu m)/F_\nu(100 \mu m)$ ratio.
The symbol coding is the same as in Fig. 7.
The cross on the top right part of the diagram indicates a typical
error bar associated with the early-type galaxies analyzed in this work
\\\\
%%%%%%%%%%%%%%%%%%%% FIG. 9 %%%%%%%%%%%%%%%%%%%%%%%%%%%%%%
\noindent
{\bf FIGURE 9}\\
The SED of the radio galaxy NGC 4486 (M87).
The points are fitted with three components: a stellar radiation field
modelled by P\'EGASE, a modified black-body law representing dust peaking
in the FIR and a power-law describing the synchrotron emission. On top of these three different
component the composite spectrum is shown. The bottom panel in each plot shows the
relative difference of the observations from the modelled spectrum.
%%%%%%%%%%%%%%%%%%%%%%%%%%%%%%%%%%%%%%%%%%%%%%%%%%%%%%%%%%
\\\\
%%%%%%%%%%%%%%%%%%%% FIG. 10 %%%%%%%%%%%%%%%%%%%%%%%%%%%%%%
\noindent
{\bf FIGURE 10}\\
Major-axis profiles of the MIR emission (6.7 $\mu m$ and 15 $\mu m$ shown
as dashed and
solid line respectively) for the edge-on lenticular galaxy NGC 5866. The emission peaks
at the nucleus and
secondary maxima at $\simeq 4$ kpc either side of the center are present.
%%%%%%%%%%%%%%%%%%%%%%%%%%%%%%%%%%%%%%%%%%%%%%%%%%%%%%%%%%

\appendix
\section{Mid-IR maps}

%%%%%%%%%%%%%%%%%% FIG. 8 %%%%%%%%%%%%%%%%%%%%%%%%%%%%%%%
\noindent
{\bf FIGURE A.1}\\
Mid-IR maps for NGC 185 (left) and NGC 205 (right).
The top panels show the $4.5 \mu m$ contours overlaid on the optical
images of the galaxies. The middle panels show the $6.7 \mu m$ maps 
and the bottom panels show the $15 \mu m$ maps with the basic contours of these maps
overlaid in order to emphasize the shape of the MIR images.
For NGC 185 the contour levels are 0.3, 0.4, 0.6, 0.8, 1.0, 1.2, 1.3, 1.4, 1.5
mJy/pixel for the 4.5 $\mu m$ map (top), 0.1, 0.2, 0.3, 0.4, 0.6, 0.8, 1.0, 1.2, 1.3, 1.4,
1.5, 1.7 mJy/pixel for the $6.7 \mu m$ map (middle) and
0.2, 0.3, 0.4, 0.6, 0.8, 1.0, 1.2 mJy/pixel for the $ 15 \mu m$ map (bottom).
For NGC 205 the contour levels are 0.4, 0.5, 0.6, 0.8, 1.0, 1.3, 1.8
mJy/pixel for the 4.5 $\mu m$ map (top), 0.2, 0.3, 0.4, 0.5, 0.6, 0.8, 1.0, 1.3, 1.7
mJy/pixel for the $6.7 \mu m$ map (middle) and 0.2, 0.3, 0.4, 0.5, 0.6, 0.8, 1.0
mJy/pixel for the $ 15 \mu m$ map (bottom).
\\\\
%%%%%%%%%%%%%%%%%% FIG. 8 %%%%%%%%%%%%%%%%%%%%%%%%%%%%%%%
{\bf FIGURE A.2}\\
Same as in Fig. A.1. but for the galaxies NGC 1052 (left) and NGC 1316 (right).
For NGC 1052 the contour levels are 0.2, 0.4, 0.6, 0.8, 1.0, 1.2, 1.4, 1.8, 2.4, 
3.0, 4.0, 5.0, 6.0, 7.0, 8.0, 9.0, 10.0, 12.0, 14.0, 16.0, 18.0
mJy/pixel for the 4.5 $\mu m$ map (top), 0.1, 0.3, 0.6, 1.0, 2.0, 4.0, 10.0, 20.0
mJy/pixel for the $6.7 \mu m$ map (middle) and 0.6, 1.0, 2.0, 4.0, 10., 20., 30., 40., 50.0
mJy/pixel for the $ 15 \mu m$ map (bottom).
For NGC 1316 the contour levels are 0.5, 1.0, 2.0, 3.0, 5.0, 8.0, 12.0, 20.0, 30.0
mJy/pixel for the 4.5 $\mu m$ map (top), 0.3, 0.6, 1.0, 1.5, 2.0, 3.0, 5.0, 10.0, 15.0
mJy/pixel for the $6.7 \mu m$ map (middle) and 0.2, 0.5, 1.0, 1.5, 2.0, 3.0, 5.0, 10.0
mJy/pixel for the $ 15 \mu m$ map (bottom).
%%%%%%%%%%%%%%%%%% FIG. 8 %%%%%%%%%%%%%%%%%%%%%%%%%%%%%%%
\\\\
{\bf FIGURE A.3}\\
Same as in Fig. A.1. but for the galaxies NGC 1399 (left) and NGC 2300 (right).
For NGC 1399 the contour levels are 0.3, 0.5, 1.0, 2.0, 4.0, 10.0, 16.0
mJy/pixel for the 4.5 $\mu m$ map (top), 0.2, 0.3, 0.5, 1.0, 2.0, 4.0, 10.0
mJy/pixel for the $6.7 \mu m$ map (middle) and 0.4, 0.6, 1.0, 2.0, 4.0, 10.0
mJy/pixel for the $ 15 \mu m$ map (bottom).
For NGC 2300 the contour levels are 0.3, 0.5, 1.0, 2.0, 4.0, 10.0
mJy/pixel for the 4.5 $\mu m$ map (top), 0.2, 0.3, 0.5, 1.0, 2.0, 4.0, 6.0
mJy/pixel for the $6.7 \mu m$ map (middle) and 0.3, 0.5, 0.7, 1.0, 1.5, 2.0
mJy/pixel for the $ 15 \mu m$ map (bottom).
%%%%%%%%%%%%%%%%%% FIG. 8 %%%%%%%%%%%%%%%%%%%%%%%%%%%%%%%
\\\\
{\bf FIGURE A.4}\\
Same as in Fig. A.1. but for the galaxies NGC 3928 (left) and NGC 4278 (right).
For NGC 3928 the contour levels are 0.5, 0.7, 1.0, 1.5, 2.0, 3.0, 4.0
mJy/pixel for the 4.5 $\mu m$ map (top), 0.3, 0.5, 1.0, 2.0, 3.0, 5.0, 10.0, 15.0, 20.0
mJy/pixel for the $6.7 \mu m$ map (middle) and 0.5, 1.0, 2.0, 3.0, 5.0, 10.0, 15.0, 20.0, 23.0
mJy/pixel for the $ 15 \mu m$ map (bottom).
For NGC 4278 the contour levels are 0.3, 0.5, 1.0, 2.0, 5.0, 10.0, 15.0
mJy/pixel for the 4.5 $\mu m$ map (top), 0.3, 0.5, 1.0, 2.0, 5.0, 10.0
mJy/pixel for the $6.7 \mu m$ map (middle) and 0.2, 0.3, 0.5, 1.0, 2.0, 4.0
mJy/pixel for the $ 15 \mu m$ map (bottom).
%%%%%%%%%%%%%%%%%% FIG. 8 %%%%%%%%%%%%%%%%%%%%%%%%%%%%%%%
\\\\
{\bf FIGURE A.5}\\
Same as in Fig. A.1. but for the galaxies NGC 4374 (left) and NGC 4473 (right).
For NGC 4374 the contour levels are 0.6, 1.0, 2.0, 5.0, 10.0, 15.0, 20.0
mJy/pixel for the 4.5 $\mu m$ map (top), 0.6, 1.0, 2.0, 5.0, 10.0, 15.0
mJy/pixel for the $6.7 \mu m$ map (middle) and 0.5, 0.7, 1.0, 1.5, 2.0, 3.0, 5.0, 6.0
mJy/pixel for the $ 15 \mu m$ map (bottom).
For NGC 4473 the contour levels are 0.7, 1.0, 1.5, 3.0, 5.0, 10.0, 15.0
mJy/pixel for the 4.5 $\mu m$ map (top), 0.3, 0.5, 0.7, 1.0, 1.5, 3.0, 5.0, 7.0, 9.0
mJy/pixel for the $6.7 \mu m$ map (middle) and 0.7, 1.0, 1.5, 3.0
mJy/pixel for the $ 15 \mu m$ map (bottom).
%%%%%%%%%%%%%%%%%% FIG. 8 %%%%%%%%%%%%%%%%%%%%%%%%%%%%%%%
\\\\
{\bf FIGURE A.6}\\
Same as in Fig. A.1. but for the galaxies NGC 4486 (left) and NGC 4649 (right).
For NGC 4486 the contour levels are 7.0, 8.0, 9.0, 10.0, 12.0, 15.0, 17.0, 19.0
mJy/pixel for the 4.5 $\mu m$ map (top), 2.0, 3.0, 4.0, 5.0, 6.0, 7.0, 8.0, 9.0, 10.0, 12.0
mJy/pixel for the $6.7 \mu m$ map (middle) and 2.0, 3.0, 4.0, 5.0, 6.0, 7.0, 8.0, 9.0
mJy/pixel for the $ 15 \mu m$ map (bottom).
For NGC 4649 the contour levels are 1.0, 1.5, 2.0, 3.0, 5.0, 8.0, 10.0, 15.0
mJy/pixel for the 4.5 $\mu m$ map (top), 0.7, 1.0, 1.5, 2.0, 3.0, 5.0, 8.0, 10.0
mJy/pixel for the $6.7 \mu m$ map (middle) and 1.0, 1.5, 2.0, 3.0, 4.0
mJy/pixel for the $ 15 \mu m$ map (bottom).
%%%%%%%%%%%%%%%%%% FIG. 8 %%%%%%%%%%%%%%%%%%%%%%%%%%%%%%%
\\\\
{\bf FIGURE A.7}\\
Same as in Fig. A.1. but for the galaxies NGC 5018 (left) and NGC 5102 (right).
For NGC 5018 the contour levels are 0.7, 1.0, 2.0, 4.0, 8.0, 15.0
mJy/pixel for the 4.5 $\mu m$ map (top), 0.5, 1.0, 2.0, 4.0, 8.0, 12.0
mJy/pixel for the $6.7 \mu m$ map (middle) and 0.5, 1.0, 2.0, 4.0, 6.0
mJy/pixel for the $ 15 \mu m$ map (bottom).
For NGC 5102 the contour levels are 0.5, 0.7, 1.0, 2.0, 4.0, 7.0
mJy/pixel for the 4.5 $\mu m$ map (top), 0.2, 0.4, 0.7, 1.0, 2.0, 4.0, 7.0
mJy/pixel for the $6.7 \mu m$ map (middle) and 0.4, 0.7, 1.0, 2.0, 4.0, 7.0
mJy/pixel for the $ 15 \mu m$ map (bottom).
%%%%%%%%%%%%%%%%%% FIG. 8 %%%%%%%%%%%%%%%%%%%%%%%%%%%%%%%
\\\\
{\bf FIGURE A.8}\\
Same as in Fig. A.1. but for the galaxies NGC 5173 (left) and NGC 5266 (right).
For NGC 5173 the contour levels are 0.3, 0.5, 1.0, 2.0, 3.0
mJy/pixel for the 4.5 $\mu m$ map (top), 0.5, 1.0, 2.0, 3.0, 3.7
mJy/pixel for the $6.7 \mu m$ map (middle) and 0.6, 1.0, 2.0, 3.0, 3.7
mJy/pixel for the $ 15 \mu m$ map (bottom).
For NGC 5266 the contour levels are 0.3, 0.5, 0.7, 1.0, 2.0, 3.0, 6.0, 10.0
mJy/pixel for the 4.5 $\mu m$ map (top), 0.5, 0.7, 1.0, 2.0, 3.0, 6.0, 10.0
mJy/pixel for the $6.7 \mu m$ map (middle) and 0.7, 1.0, 2.0, 3.0, 4.0, 6.0, 8.0
mJy/pixel for the $ 15 \mu m$ map (bottom).
%%%%%%%%%%%%%%%%%% FIG. 8 %%%%%%%%%%%%%%%%%%%%%%%%%%%%%%%
\\\\
{\bf FIGURE A.9}\\
Same as in Fig. A.1. but for the galaxies NGC 5363 (left) and NGC 5866 (right).
For NGC 5363 the contour levels are 1.0, 1.5, 2.0, 3.0, 5.0, 10.0, 13.0
mJy/pixel for the 4.5 $\mu m$ map (top), 0.7, 1.0, 1.5, 2.0, 3.0, 5.0, 10.0, 15.0
mJy/pixel for the $6.7 \mu m$ map (middle) and 0.8, 2.0, 5.0, 10.0
mJy/pixel for the $ 15 \mu m$ map (bottom).
For NGC 5866 the contour levels are 0.5, 1.0, 2.0, 4.0, 8.0, 10.0
mJy/pixel for the 4.5 $\mu m$ map (top), 0.3, 0.5, 1.0, 2.0, 4.0, 8.0, 12.0
mJy/pixel for the $6.7 \mu m$ map (middle) and 0.5, 1.0, 2.0, 4.0, 6.0, 8.0, 12.0
mJy/pixel for the $ 15 \mu m$ map (bottom).
%%%%%%%%%%%%%%%%%%%%%%%%%%%%%%%%%%%%%%%%%%%%%%%%%%%%%%%%%

\end{document}